\documentclass{nature}
\usepackage{epsf}

\def\AU{{\sc au}}
\def\bfAU{{\footnotesize\bf AU}}
\def\cr105{2000~CR$_{105}$}
\def\vb12{2003~VB12}
\def\qflyby{q_f}
\def\iflyby{i_f}

\title{Stellar encounters as the origin of distant solar system objects 
in highly eccentric orbits}

\author {Scott J. Kenyon$^1$ \& Benjamin C. Bromley$^2$} 

\begin{document}

\spacing{1}

\maketitle

\begin{affiliations}
\item Smithsonian Astrophysical Observatory, 60 Garden Street,
Cambridge, MA 02138, USA
\item Department of Physics, University of Utah, 201 JFB, Salt Lake City,
UT 84112, USA
\end{affiliations}

\begin{abstract}

The Kuiper Belt\cite{luu02} extends from the orbit of Neptune at 30 \bfAU\ 
to an abrupt outer edge at $\sim$ 50~\bfAU\ from the Sun\cite{all01}. 
Beyond the edge is a sparse population of objects with large orbital 
eccentricities\cite{tru00,gla01}. Neptune shapes the dynamics of these 
objects, but the recently discovered planet \vb12\ (Sedna\cite{bro04})
has an eccentric orbit with a perihelion distance, 70~\bfAU, far beyond 
Neptune's gravitational influence\cite{lev97,gla02,lev03}.  Although 
influences from passing stars could have created the Kuiper Belt's outer 
edge and could have scattered objects into large, eccentric orbits\cite{fer00,ida00}, 
no model currently explains the properties of Sedna. Here we show that a 
passing star probably scattered Sedna from the Kuiper Belt into its observed 
orbit.  The likelihood that a planet at 60--80 \bfAU\ can be scattered 
into Sedna's orbit is $\sim$ 50\%; this estimate depends critically on 
the geometry of the flyby.  Even more interesting, though, is the $\sim$ 
10\% chance that Sedna was captured from the outer disk of the passing star.  
Most captures have very high inclination orbits; detection of these objects
would confirm the presence of extrasolar planets in our own Solar System.

\end{abstract}

In the planetesimal theory, planets grow from mergers of smaller,
slowly-moving objects embedded in a gaseous, circumstellar disk 
surrounding a young star\cite{saf69}.  For reasonable initial disk 
masses\cite{wya03}, this process yields\cite{ken99,ken02} 1000~km 
planets with roughly circular orbits in 50--100 Myr at 70~\AU. To scatter 
a 1000~km planet into a Sedna-like orbit, a much larger planet must 
form at roughly the same distance.  These more massive planets take 
at least 1 Gyr to form and should be rare.  Despite extensive searches, 
none have been detected\cite{bro04,bru02}.  Thus, {\it in situ} 
formation of a planet with Sedna's current properties seems unlikely.

A chance encounter between the Sun and another star is a more 
plausible explanation for the orbit of Sedna.  Previous numerical
simulations show that a flyby of a solar-type star, with a distance 
of closest approach $\qflyby \sim$~150--200~\AU, scatters objects at
60--80~\AU\ into very eccentric orbits and truncates the Kuiper Belt
at its observed edge\cite{fer00,ida00}.  A more distant flyby, with 
$\qflyby \sim$ 500--1000~\AU, lifts objects from Neptune crossing-orbits 
into Sedna-like orbits, leaving the rest of the Kuiper Belt relatively 
unchanged\cite{mor04}.  Although other mechanisms to truncate the Kuiper 
Belt are possible\cite{wei03,you03,ada04}, they are inconsistent with 
the apparent formation of 1000~km planets in the larger disks observed 
around several nearby stars\cite{gre98,gre04}.  Here we improve on previous
flyby calculations and derive an initial orbital distribution of Kuiper
belt objects 
from a detailed planet formation calculation\cite{kb04a,kb04b}. We then
use a complete set of N-body simulations to provide the first estimates
of the likelihood that a close encounter yields the observed edge to 
the Kuiper Belt and planets in Sedna-like orbits.

The discovery of Sedna also prompted us to consider an exciting new
possibility, captures from the Kuiper Belt of the passing star\cite{mor04}.
Like most stars, the Sun probably formed in a star cluster and experienced 
a close encounter with another cluster member\cite{ada01}. Encounters 
with field stars are unlikely\cite{gar99}. Cluster lifetimes of 100~Myr 
to 1~Gyr\cite{ada01} allow a solar-type star to produce 1,000~km objects 
as far out as 100~\AU.  Thus, both the Sun and the passing star probably 
had extended planetary disks at the start of their encounter.  By analogy 
with galaxy collisions\cite{too72,bar99}, the two stars can exchange a 
significant amount of outlying material during a flyby, depending on the 
geometry of the encounter and the properties of the planetary systems.  
Here we adopt a cluster age of 30--200 Myr for the encounter, and use 
planet formation and N-body simulations to provide the first assessment 
of the probability that a close pass yields the observed edge to the 
Kuiper Belt and the capture of an extrasolar planet into a Sedna-like orbit.

Consistent with formation in a star cluster with a velocity 
dispersion\cite{gir89} of 1~km~s$^{-1}$, we assume the star and
the Sun have equal mass and orbit as a marginally bound pair. Modest
differences in starting conditions have no qualitative impact on our 
results. During the flyby, tidal shear dramatically increases orbital 
eccentricities in the disk at large heliocentric distances; objects 
closer to the Sun are unperturbed.  Simulations that produce a sharp 
edge in the Kuiper Belt yield a correlation between $q_f$ and $i_f$, 
the inclination of the flyby trajectory relative to the invariable 
plane of the Solar System.
For low-inclination orbits, where the passing star corotates with the 
disk, $q_f$ must exceed 120~\AU\ to avoid disrupting the Kuiper Belt. 
Counterrotating flybys with $i_f$ = 90$^\circ$--180$^\circ$ require 
smaller $q_f$ to produce the observed edge. The simulations suggest 
$i_f(\mbox{\rm in~degrees}) \sim 275 - 1.6 q_f$.  For a specific 
$q_f$ =90--160 \AU, this expression yields $i_f$ to $\sim$ 10\% for 
corotating encounters and to $\sim$ 30\% for counterrotating orbits, 
where tidal shear is more broadly distributed in the disk and the 
edge is less sharp.  
In both cases, post-flyby collisional grinding removes additional 
material from the Kuiper Belt and accentuates the outer edge.  
Detailed collision models\cite{kb04a,kb04b} suggest that grinding 
and shear reduce the mass by 98\% to 99.9\%, close to observational 
estimates\cite{ber04}.

We may further constrain the passing star's trajectory by assuming
that Sedna is indigenous to the Solar System. Simulations then identify
the flyby configurations that scatter objects into Sedna-like orbits,
defined to have modest inclination ($i<30^\circ$), and large
eccentricity and perihelion distance ($e>0.5$ and $q>50$~\AU).  Fig.~2
shows results for two encounters where planets are scattered from 
nearly circular orbits at 80~\AU\ into Sedna-like orbits, a corotating 
flyby with $\qflyby = 160$~\AU\ and $\iflyby = 23^\circ$, and a 
counterrotating flyby with $\qflyby = 90$~\AU\ and $\iflyby = 172^\circ$. 
Both flybys scatter $\sim$ 10\% of the objects from an annulus at 
80$\pm$2.5~\AU\ into high eccentricity orbits. After 50--200 Myr, 
our coagulation calculations produce 5--25 planets with radii of 
500--2000~km at 80$\pm$2.5 \AU. Thus, a flyby that makes the current 
edge to the Kuiper Belt also produces at least one indigenous object 
in a Sedna-like orbit.  Roughly 95\% of these `successful' flybys are 
corotating encounters.  The chance of a Sedna-like orbit is fairly 
independent of the starting distance or the initial orbital eccentricity. 
Roughly 2\% of the 40--200 large planets formed in circular orbits at 
60--80 \AU\ scatter into a Sedna-like orbit.  We derive similar results 
for the 50--500 planets with $e > 0.5$ and $q >$ 35 \AU\ in a 
`scattered disk' (Fig.~3), suggesting that formation of a Sedna-like 
orbit from a scattered disk is 2--3 times more likely than formation 
from planets in initially circular orbits.  Assuming both paths produce 
Sedna-like orbits, we predict 3--30 Sedna-like objects compared to the 
30--100 estimated from the detection of a single Sedna at 90 \AU.

Surprisingly, the probability that captured planets have Sedna-like orbits 
is also significant. When the trajectory of the Sun and the disk of the 
passing star corotate, the Sun captures up to a third of the extrasolar 
objects initially in orbit at distances of 60--80~\AU\ from the passing 
star (Fig.~2).  This `capture efficiency' depends on $q_f$ and the 
angle $i_f^{\prime}$ between the rotational axes of the two disks.  
The capture efficiency is 10\% to 30\% for $\qflyby \le 120$~\AU\ and 
$i_f^{\prime} \le 80^\circ$, and falls to $\sim$ 5\% for $\qflyby \sim$ 
160~\AU\ and $i_f^{\prime} \le 45^\circ$.  For flybys with many captures, 
counterrotating flybys yield more Sedna-like objects than corotating flybys. 
For a random ensemble of flybys that produce the observed edge of the Kuiper 
Belt, the probability of a capture into a Sedna-like orbit is $\sim$ 5\%.

The probability estimates for indigenous and captured planets with
Sedna-like orbits assume a flyby that produces the observed outer edge 
of the Kuiper Belt.  Simple estimates of the evolution of star clusters 
indicate a reasonable probability, $\sim$ 15\%, for a stellar encounter 
within 160~\AU\ during the first 1 Gyr of the solar lifetime\cite{ada01}. 
Because the Sun is heavy enough to remain in the cluster as the cluster 
evaporates, 30\% to 50\% of the encounters probably occur during the first 
100 Myr of the solar lifetime.  Thus the total probability to produce at 
least one object with a Sedna-like orbit in the Solar System is reasonably 
large, $\sim$ 5\% to 10\% for an indigenous object and $\sim$ 1\% for a 
captured object.  These probabilities are significant compared to the chance 
of finding a Solar System like our own around any solar-type star, $\sim$ 
1\% or less\cite{ada01}.

The simulations demonstrate that flybys leave unique signatures on the 
dynamics of the outer solar system. Corotating flybys produce a single 
population with orbital inclination $i \sim 10^\circ$ at 100--500~\AU; 
counterrotating flybys produce a cold population with $i \le 10^\circ$ 
and a hotter population with $i \ge 30^\circ$ (Fig.~2).  Flybys with an 
initial scattered disk of objects at 60--80~\AU\ yield broader, but still 
distinct, distributions in inclination (Fig.~3).  The two inclination 
populations of counterrotating flybys qualitatively resemble the two 
observed populations of Kuiper belt objects at 40--50~\AU\cite{bro01}.  
Systematic 
searches for high inclination objects in the outer Solar System can 
distinguish between corotating and counterrotating flybys and can 
estimate the relative fraction of objects in a scattered disk at 
the time of the flyby.

Orbits of individual objects also test the models.  Some flybys produce 
objects with orbital elements similar to \cr105, a $\sim$ 1,000~km 
planet with $q = 44$~\AU, $e = 0.8$, and $i = 22^\circ$. Although 
\cr105\ is closer to Neptune than Sedna, the known planets probably 
cannot scatter \cr105\ into its present orbit\cite{gla02,mor04}.  
The corotating flyby in Fig.~2 cannot produce orbits with the large 
range of inclination observed in Sedna and \cr105, but counterrotating 
flybys yield a broad range encompassing the observations.  Both types 
of flyby generate orbits similar to \cr105\ and Sedna from a scattered 
disk (Fig.~3; ref. 16).  Long term simulations suggest that interactions with 
Neptune broaden the inclination distributions of corotating flybys, 
forming objects with orbital elements closer to both \cr105\ and Sedna.  
From these simulations, we estimate that \cr105\ is 2--3 times more likely 
to be a captured planet than Sedna.  Because our calculations are the 
only known way to produce high inclination objects, searches at high
ecliptic latitude provide the best test of this picture for Sedna formation.
Detection of objects with $i \ge 40^{\circ}$ would clinch the case for 
the presence of extrasolar planets in the outer Solar System.




\begin{addendum}
 \item We acknowledge a generous allotment, $\sim$ 3000 cpu days,
of computer time at the supercomputing center at the Jet Propulsion 
Laboratory through funding from the NASA Offices of Mission to Planet 
Earth, Aeronautics, and Space Science.
Advice and comments from M. Geller and two anonymous referees 
improved our presentation.  
The {\it NASA} {\it Astrophysics Theory Program} supported part 
of this project.
 \item[Competing Interests] The authors declare that they have no
competing financial interests.
 \item[Correspondence] Correspondence and requests for materials
should be addressed to \\
S.J.K. (skenyon@cfa.harvard.edu).
\end{addendum}

\clearpage


\begin{figure}
\epsfxsize 7.0in
\hskip -7ex
\epsffile{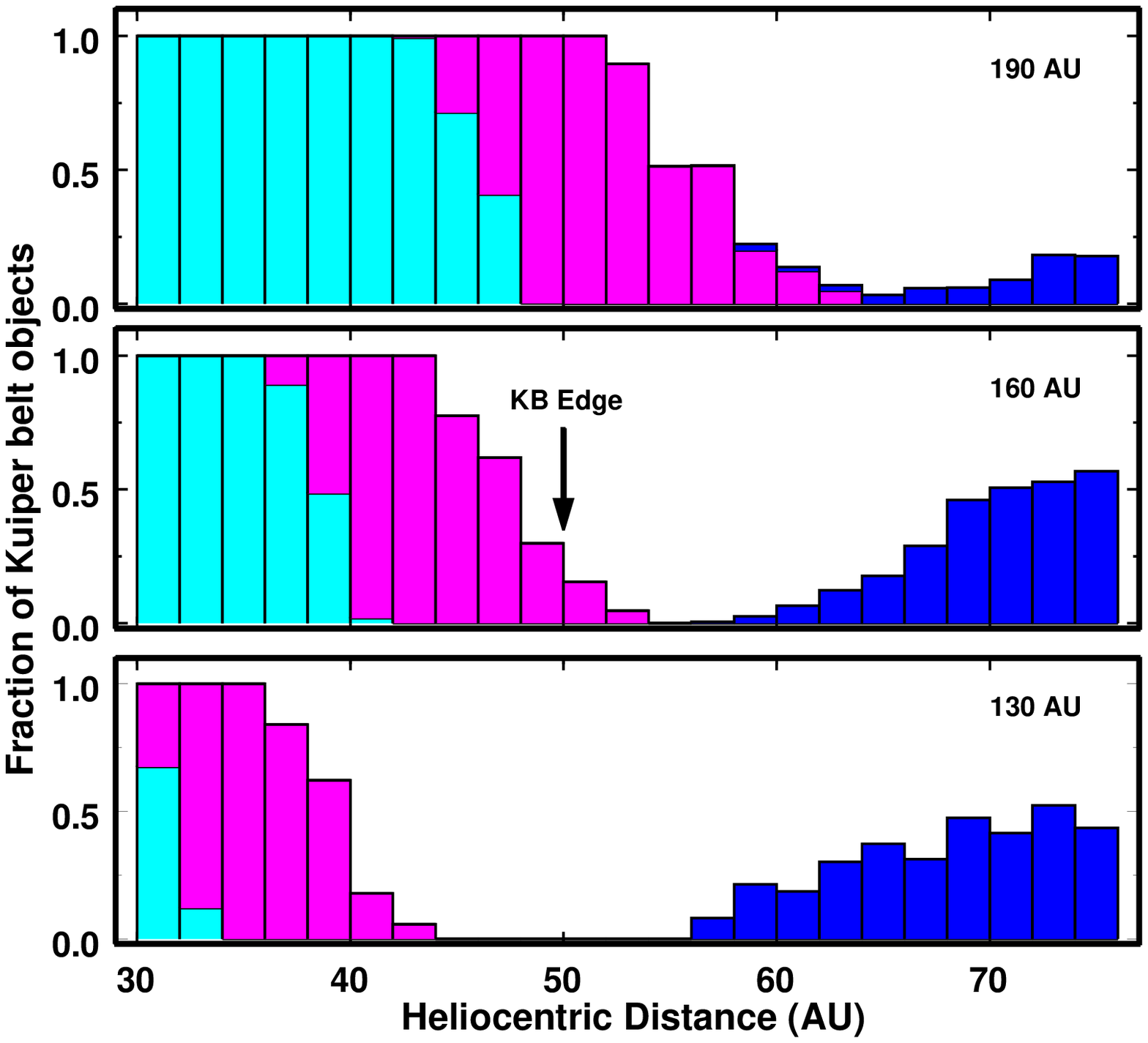}
\end{figure}
\noindent
{{\bf Figure 1} Stirring of the eccentricities of planets by the close pass 
of a Sun-like star.  The star is on a marginally bound orbit corotating
with the Sun's disk, with an orbital inclination of $i=23^\circ$ and
argument of perihelion $\omega=212^\circ$.  Each panel lists the distance 
of closest approach.  The histograms show the frequency of the final 
eccentricity, $e_f$, for 20,000 particles initially in circular orbits 
at 40--80 \AU\ around the Sun.  Magenta histograms: fraction of
orbits with $e_f<$0.04; cyan histograms: fraction of orbits with
$e_f<0.2$; blue histograms: fraction of orbits with Sedna-like orbits
($e_f>0.5$).  The 160 \AU\ encounter places the edge of the Kuiper Belt
where observed, as indicated by the arrow in the middle panel.}

\begin{figure}
\epsfxsize 6.5in
\hskip -9ex
\epsffile{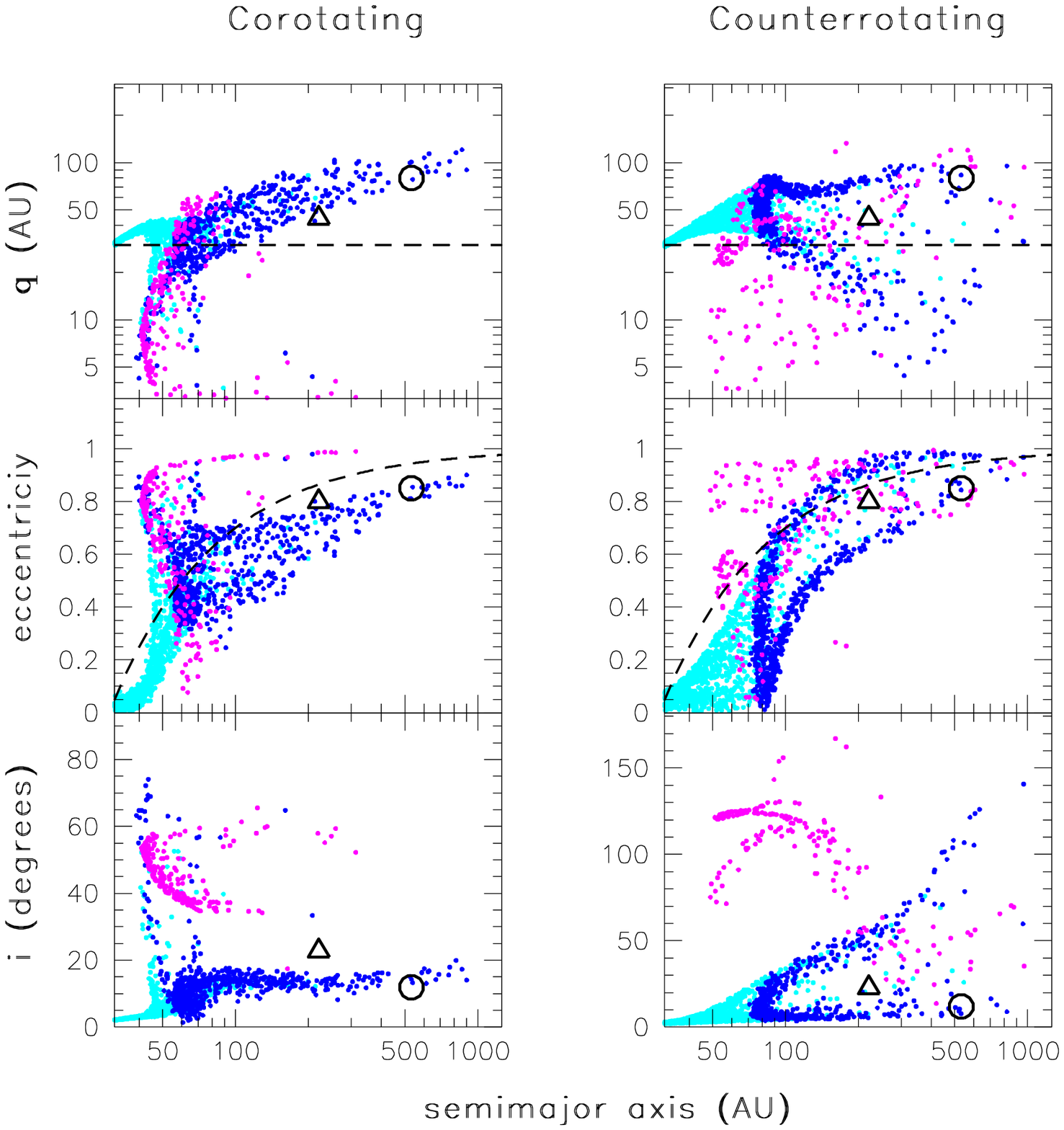}
\end{figure}
\vskip -10ex
\noindent
{{\bf Figure 2} Orbital elements of planets, formed {\it in situ} 
in the planetary disk, after the flyby of a 1~solar mass star as 
a function of final heliocentric distance. The planets have an
initial probability $p$ of formation at in 1 AU bins at 40--80 \AU\ 
derived from detailed planet formation calculations, $ p \propto a^{-1}$, 
where $a$ is the semimajor axis.
Left panels: a marginally bound, corotating flyby
($d_{close}$ = 160~\AU, $i=23^\circ$, and $\omega=212^\circ$).
Right panels: a marginally bound counterrotating flyby 
($d_{close}=90$~\AU, $i=172^\circ$ degrees, and $omega=130^\circ$). 
Large black circles: orbital elements of Sedna.
Large black triangles: orbital elements of \cr105.
In the upper panels, objects below the dashed lines cross 
the orbit of Neptune; in the middle panels, objects
above the dashed lines cross the orbit of Neptune.
Cyan points: planetary orbits initially distributed 
between 30--80 \AU\ (Fig. 2), with initial eccentricity $e_0 =$ 0.02
and initial inclination $i_0$ = 0.5$^\circ$.
Blue points: orbits initially at $80\pm 2.5$~\AU\ with $e_0$ = 0.05
and $i_0$ = 1$^\circ$.
Magenta points: objects captured by the Sun from the disk of the 
passing star.  Each simulation consisted of 1,000 particles.

}

\begin{figure}
\epsfxsize 7.0in
\hskip -9ex
\epsffile{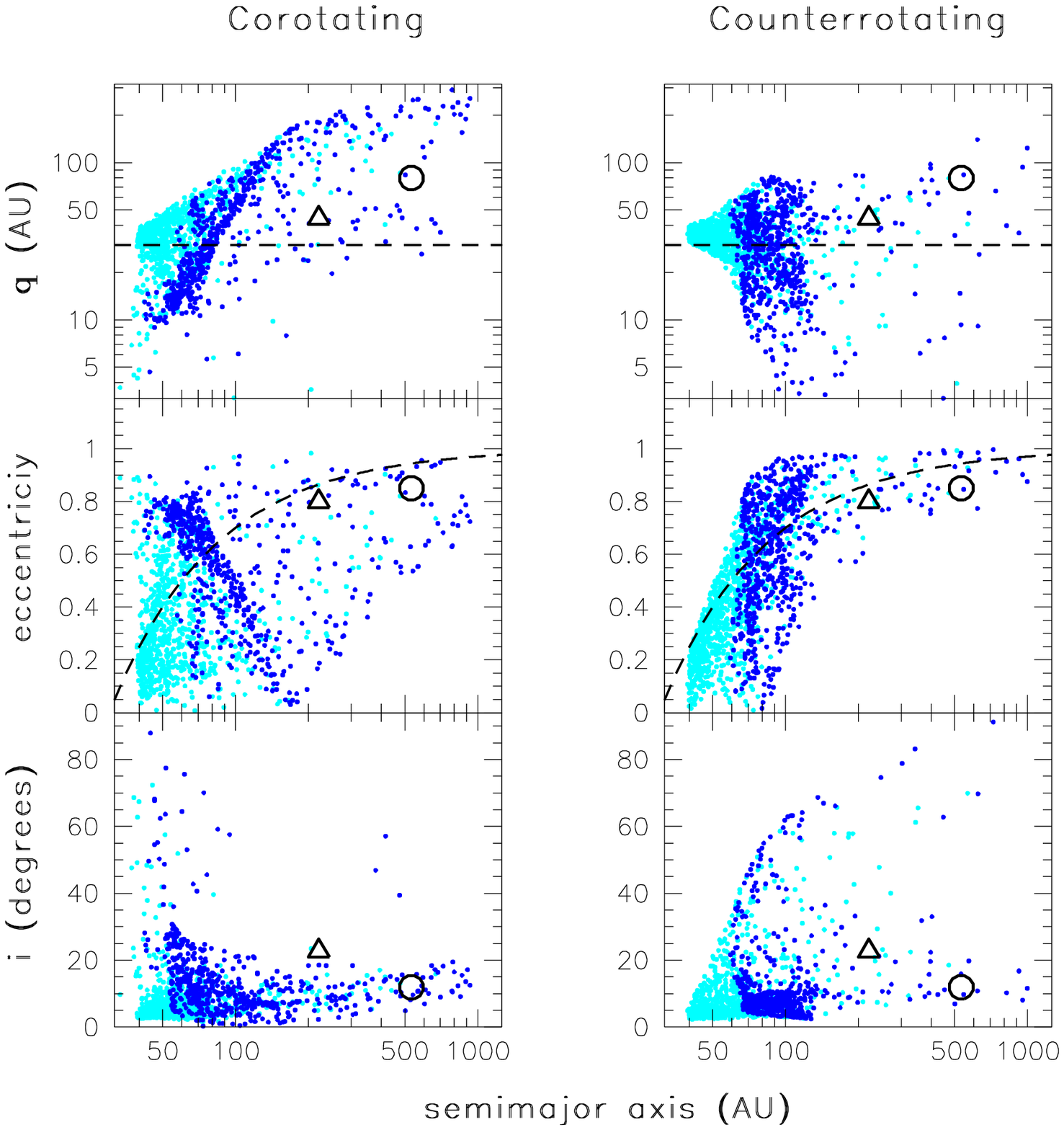}
\end{figure}
\vskip -8ex
\noindent
{{\bf Figure 3} Orbital elements of planets, initially in a `scattered
disk,' after a flyby, as described in Fig.~2.  The planets are placed 
initially at 40--80 \AU\, with $p \propto a^{-1}$ as in Fig. 2,
$q$ = 35 AU, and $i_0 \le$ 5$^\circ$.  Cyan points: planetary orbits 
initially distributed between 60--80~\AU\ with perihelion distance 
of 35~\AU.  Blue points: orbits initially at $80\pm 2.5$~\AU,
also with perihelion distance of 35~\AU.  The objects in the figure
end up with a broader range of orbital elements than those in
Fig.~2. While a subset of particles are consistent with both the
orbits of Sedna and \cr105, the distributions are broader, 
reducing the correlations between the orbital elements of scattered
objects and the good agreement between the observations and the models 
compared with the case of {\it in situ} formation shown in Fig.~2.  }


\vskip 3ex

\begin{thebibliography}{99}

\bibitem{luu02} Luu, J.~X.~\& Jewitt, 
D.~C.\ Kuiper Belt Objects: Relics from the Accretion Disk of the Sun.\  
{\it Ann. Rev. Astron. Astrophys.} {\bf 40}, 63-101 (2002).

\bibitem{all01} Allen, R. L., Bernstein, G. M., \& Malhotra, R. 
The edge of the solar system, {\it Astrophys. J.} {\bf 549}, L241-L244 (2001)

\bibitem{tru00} Trujillo, C., Jewitt, D., \& Luu, J.,
Population of the scattered Kuiper belt.\
{\it Astrophys. J.} {\bf 102}, 529-533 (2002).

\bibitem{gla01} Gladman, B., Kavelaars, 
J.~J., Petit, J., Morbidelli, A., Holman, M.~J., \& Loredo, T.\ The 
Structure of the Kuiper Belt: Size Distribution and Radial Extent.\  
{\it Astron. J.},{\bf 122}, 1051-1066 (2001).

\bibitem{bro04} Brown, M. E., Trujillo, C. \&  Rabinowitz, D. 
Discovery of a candidate inner Oort cloud planetoid.\ astro-ph/0404456 (2004).

\bibitem{lev97} Levison, H. F., \& Duncan, M. J. From the Kuiper belt
to Jupiter-family comets: The spatial distribution of of ecliptic comets.\
{\it Icarus} {\bf 127}, 13-32 (1997).

\bibitem{gla02} Gladman, B., Holman, 
M., Grav, T., Kavelaars, J., Nicholson, P., Aksnes, K., \& Petit, J.-M.\ 
Evidence for an Extended Scattered Disk.\ {\it Icarus} {\bf 157}, 269-279 
(2002).

\bibitem{lev03} Levison, H., \& Morbidelli, A. The formation of the
Kuiper belt by the outward transport of bodies during Neptune's 
migration.\ {\it Nature} {\bf 426}, 419-421 (2003).

\bibitem{fer00} Fern{\' a}ndez, J.~A. \& Brunini, A.  The buildup
of a tightly bound comet cloud around an early Sun immersed in a dense
Galactic environment: Numerical experiments.\ {\it Icarus} {\bf 106},
580-590 (2000).

\bibitem{ida00} Ida, S., Larwood, J., \& Burkert, A. Evidence for early
stellar encounters in the orbital distribution of Edgeworth-Kuiper belt
objects.\ {\it Astrophys. J.} {\bf 528}, 351-356 (2000).


\bibitem{saf69} Safronov, V. S. Evolution of the protoplanetary cloud 
and formation of the earth and planets.\ Nauka, Moscow (1969).
[Translation 1972, NASA TT F-677]

\bibitem{wya03} Wyatt, M. C., Dent, W. R. F., \& Greaves, J. S.
SCUBA observations of dust around Lindroos stars: evidence for a 
substantial submillimetre disc population.\ {\it Mon. Not. Roy. Astr. Soc.}
{\bf 342,} 876-888 (2003).


\bibitem{ken99} Kenyon, S. J., Wood, K., Whitney, B. A., \& Wolff, M. 
Forming the Dusty Ring in HR 4796A.\ {\it Astrophys. J.}, {\bf 524}, 
L119-L123 (1999).


\bibitem{ken02} Kenyon, S. J. Planet formation in the outer solar system.\ 
{\it Pub. Astron. Soc. Pac.} {\bf 114}, 265-283 (2002).

\bibitem{bru02} Brunini, A.~\& 
Melita, M.~D.\ The Existence of a Planet beyond 50 AU and the Orbital 
Distribution of the Classical Edgeworth-Kuiper-Belt Objects.\  {\it 
Icarus},{\bf 160}, 32-43 (2002). 

\bibitem{mor04} Morbidelli, A., \& Levison, H. Scenarios for the Origin of 
the Orbits of the Trans-Neptunian Objects 2000 CR105 and \vb12. \ 
{\it Astron. J.},{\bf 128}, 2564-2576 (2004). 

\bibitem{wei03} Weidenschilling, S. J. Radial drift of particles in the 
solar nebula: implications for planetesimal formation.\ {\it Icarus}
{\bf 165}, 438-442 (2003).

\bibitem{you03} Youdin, A. N., \& Shu, F. H. Planetesimal Formation by 
Gravitational Instability.\ {\it Astrophys. J.} {\bf 580}, 494-505 (2003).

\bibitem{ada04} Adams, F. C., Hollenbach, D., Laughlin, G., \& Gorti, U.
Photoevaporation of Circumstellar Disks Due to External Far-Ultraviolet 
Radiation in Stellar Aggregates.\ {\it Astrophys. J.} {\bf 611}, 360-379 (2004).

\bibitem{gre98} Greaves, J. S., Holland, W. S., Moriarty-Schieven, G., 
Jenness, T., Dent, W. R. F., Zuckerman, B., McCarthy, C., Webb, R. A., 
Butner, H. M., Gear, W. K., Walker, H. J.  A Dust Ring around $\epsilon$ 
Eridani: Analog to the Young Solar System. 
{\it it Astrophys. J.} {\bf 506} L133-L137.

\bibitem{gre04} Greaves, J. S., Wyatt, M. C., Holland, W. S., \& Dent, 
W. R. F.  The debris disc around $\tau$ Ceti: a massive analogue to 
the Kuiper Belt.\ {\it Mon. Not. Roy. Astron. Soc.} {\bf 351}, L54-L58 (2004).

\bibitem{kb04a} Kenyon, S. J., \& Bromley, B. C. Collisional cascades 
in planetesimal disks. II. embedded planets.\ {\it Astron. J.} {\bf 127}, 
513-530 (2004).

\bibitem{kb04b} Kenyon, S. J., \& Bromley, B. C. The size distribution
of Kuiper belt objects. {\it Astron. J.} {\bf 128}, no. 4 [astro-ph/0406556] (2004).

\bibitem{ada01} Adams, F. C., \& Laughlin, G. 
Constraints on the birth aggregate of the Solar System. \
{\it Icarus} {\bf 150}, 151-162 (2001).

\bibitem{gar99} Garcia-Sanchez, J., Preston, R. A., Jones, D. L.,
Weissman, P. R., Lestrade, J.-F., Latham, D. W., \& Stefanik, R. P.
Stellar encounters with the Oort cloud based on HIPPARCOS data.\
{\it Astron. J.} {\bf 117}, 1042-1055 (1999).

\bibitem{too72} Toomre, A., \& Toomre, J. Galactic Bridges and Tails. \
{\it Astrophys. J.} {\bf 178}, 623-666 (1972).

\bibitem{bar99} Barton, E. J., Bromley, B. C., \& Geller, M. J.
Kinematic Effects of Tidal Interaction on Galaxy Rotation Curves. \
{\it Astrophys. J.} {\bf 511}, L25-L28 (1999).

\bibitem{gir89} Girard, T. M., Grundy, W. M., Lopez, C. E., 
\& van Altena, W. F. Relative proper motions and the stellar 
velocity dispersion of the open cluster M67.\ {\it Astron. J.}
{\bf 98}, 227-243 (1989).

\bibitem{ber04} Bernstein, G. M., Trilling, D. E., Allen, R. L., Brown, M. E.,
Holman, M. J., \& Malhotra, R., The Size Distribution of Trans-Neptunian 
Bodies.\ {\it Astron. J.},{\bf 128}, 1364-1390 (2004).

\bibitem{bro01} Brown, M. E. The inclination distribution of the Kuiper belt.\
{\it Astron. J.} {\bf 121}, 2804-2814 (2001).














\end{thebibliography}
\end{document}

The requirement that a stellar flyby produces the Kuiper belt's outer
edge helps to narrow down the possible binary orbital elements. 
We identify an edge-producing if it can yield planets with
$e < 0.04$ inside 30~\AU, and $e>0.2$ outside 50~\AU.  In our
simulations these constraints yield a correlation between the
inclination of the passing star, $i_{\rm binary}$, and the distance of
closest approach, $d_{\rm close}$.  For example, if $i_{\rm binary}
\sim 20^\circ$, then the optimal $d_{\rm close}$ is near $160$~\AU
(Fig.~1). If the perihelion were significantly smaller, the Kuiper
Belt would be totally disrupted, while a larger value of $d_{\rm
close}$ would not truncate the disk. At higher inclination the optimal
distance of closest approach is smaller, balancing the reduced time
for interaction between passing star and planets with an increase in
the star's gravitational influence. We find that for a given binary
inclination, the relation $d_{\rm close}/\mbox{\rm \AU} = 0.62 i_{\rm
binary}/{\rm degree} + 170$ approximately gives the optimal distance
of closest approach in the range 90--160~\AU. In corotating flyby
encounters ($i_{\rm binary}<90^\circ$) $d_{\rm close}$ must be within
$\sim$10\% of this estimate.  
